\documentclass[superscriptaddress,reprint,secnumarabic,amssymb, nobibnotes, aps, pra]{revtex4-1}

\DeclareMathAlphabet{\mathpzc}{OT1}{pzc}{m}{it}

\usepackage{graphics}
\usepackage{epsfig}
\usepackage{epstopdf}
\usepackage{subfigure}
\usepackage{amsmath}
\usepackage{subfigure}
\usepackage{float}
\usepackage{enumerate}
\usepackage{array}
\usepackage{pifont}
\usepackage[11pt]{moresize}
\usepackage{datetime}

\usepackage{mathrsfs}
\newcolumntype{C}[1]{>{\centering\let\newline\\\arraybackslash\hspace{0pt}}m{#1}}
\newcolumntype{N}{@{}m{0pt}@{}}

\begin{document}
	
	%\preprint{}
	
	\title{$\mathpzc{PT}$-symmetric spectral singularity and negative frequency resonance}
	
	

	\author{Sarang Pendharker}
	\affiliation{Department of Electrical and Computer Engineering, University of Alberta, Edmonton, Alberta T6G 1H9, Canada}
	\author{Yu Guo}
	\affiliation{Department of Electrical and Computer Engineering, University of Alberta, Edmonton, Alberta T6G 1H9, Canada}
	\author{Farhad Khosravi}
	\affiliation{Department of Electrical and Computer Engineering, University of Alberta, Edmonton, Alberta T6G 1H9, Canada}
	\author{Zubin Jacob}
	\email[]{zjacob@ualberta.ca}
	\affiliation{Department of Electrical and Computer Engineering, University of Alberta, Edmonton, Alberta T6G 1H9, Canada}
	\affiliation{Birck Nanotechnology Center, School of Electrical and Computer Engineering, Purdue University, West Lafayette, IN 47906, USA }
	\email[]{zjacob@ualberta.ca}
	%\thanks{}
	
	
	\date{\today}
	
	\begin{abstract}
		Vacuum consists of a bath of balanced and symmetric positive and negative frequency fluctuations. Media in relative motion or accelerated observers can break this symmetry and preferentially amplify negative frequency modes as in Quantum Cherenkov radiation and Unruh radiation. Here, we show the existence of a universal negative frequency-momentum mirror symmetry in the relativistic Lorentzian transformation for electromagnetic waves. We show the connection of our discovered symmetry to parity-time ($\mathpzc{PT}$) symmetry in moving media and the resulting spectral singularity in vacuum fluctuation related effects.   We prove that this spectral singularity can occur in the case of two metallic plates in relative motion interacting through positive and negative frequency plasmonic fluctuations (negative frequency resonance). Our work paves the way for understanding the role of $\mathpzc{PT}$-symmetric spectral singularities in amplifying fluctuations and motivates the search for $\mathpzc{PT}$-symmetry  in novel photonic systems.
		
	\end{abstract}
	
	%\keywords{}
	
	\maketitle

	\section{Introduction}

Systems with $\mathpzc{PT}$-symmetric Hamiltonians have invoked interest in recent years, primarily because they enable the extension of quantum mechanical formulation to systems with complex non-Hermitian Hamiltonians \cite{bender1998real}. Bender et. al. discovered that \cite{bender2005introduction,bender2007making} that for an energy eigen-spectrum to be real, the stringent condition of Hermiticity of a Hamiltonian can be replaced by a weaker $\mathpzc{PT}$-symmetry condition. A major consequence of this extension of quantum mechanical framework to non-Hermitian systems, is a new class of optical structures \cite{ruter2010observation} with spatially distributed loss and gain profiles \cite{el2007theory,makris2008beam,guo2009observation}. Such $\mathpzc{PT}$-symmetric non-Hermitian optical systems with complex dielectric profiles find promising applications in optical components ranging from couplers \cite{principe2015supersymmetry} and waveguides \cite{alaeian2014non} to microresonators \cite{chang2014parity,hassan2016integrable} and lasers \cite{feng2014single,hodaei2014parity,hodaei2015parity,hodaei2016single,hodaei2016design,longhi2010pt,chong2011p}.
%such as plasmonic waveguides \cite{alaeian2014non}, non-hermitian optical couplers \cite{principe2015supersymmetry}, $\mathpzc{PT}$-symmetric optical switches \cite{lupu2013switching,lupu2014using}, nonlinear optical oscillators \cite{hassan2016integrable}, single mode lasers \cite{feng2014single,hodaei2014parity,hodaei2015parity,hodaei2016single,hodaei2016design}, simultaneous lasing and coherent perfect absorption \cite{longhi2010pt,chong2011p}, $\mathpzc{PT}$-symmetric micro-resonators \cite{chang2014parity} and active electronic circuits \cite{schindler2011experimental,ramezani2012bypassing}. 
	
An important characteristic of the $\mathpzc{PT}$-symmetric systems is that they exhibit spectral singularities (zero-width resonance) \cite{ahmed2009zero,mostafazadeh2009spectral}. The $\mathpzc{PT}$-symmetric spectral singularities have been observed in a variety of systems such as, periodic finite gap systems \cite{correa2012spectral}, confined optical potential \cite{sinha2013spectral}, and unidirectional singularities in Fano coupled disk resonators \cite{ramezani2014unidirectional}. Recently, the $\mathpzc{PT}$-symmetric singularity in a graphene metasurface has been employed for enhanced sensing applications \cite{chen2016p}. The stabilities and instabilities in a complex potential system are also related to $\mathpzc{PT}$-symmetry \cite{kirillov2012pt}. Therefore characterization of the $\mathpzc{PT}$-symmetry in a complex Hamiltonian system is important not just to enable consistent quantum mechanical formulation, but also to identify the stable and unstable regimes in photonic systems, and to predict singularities. 
	
A moving lossy medium such as a plasma in motion is known to exhibit electromagnetic instabilities \cite{silveirinha2014optical}. These instabilities in a moving medium are caused by the Cherenkov amplification of negative energy waves \cite{nezlin1976negative} and have been recently linked to the noncontact vacuum friction \cite{pendry1997shearing,pendry2010quantum,volokitin2011quantum,vcPhysRevA.88.042509} between media at relative motion. Vacuum friction arises from quantum-fluctuation induced near-field photonic interactions \cite{intravaia2014quantum}, and has also been studied in particles moving near surfaces \cite{dedkov2009fluctuation,intravaia2011fluctuation,pieplow2015cherenkov,intravaia2015friction} and in rotating bodies \cite{manjavacas2010vacuum,dedkov2012fluctuation}. The nature of vacuum friction is to oppose the relative motion, and therefore the energy spent in maintaining the relative velocities is utilized in the amplification of vacuum fluctuations, which results in the instabilities. Recently, Silveirinha \cite{silveirinha2014optical,PhysRevA.94.033810,silveirinha2014spontaneous} reported that these instabilities in moving media occur because of $\mathpzc{PT}$-symmetry breaking. Recently, Guo et. al. \cite{guo2014singular} have shown that moving media can support singular resonances, which are manifested in giant vacuum friction and enhanced non-equilibrium heat transfer between two moving slabs \cite{guo2014giant}. However, the origin of these singular resonances in view of the symmetries present in moving media remains unexplained. 
	
In this paper, we reveal a $\mathpzc{PT}$-symmetric spectral singularity (zero width resonance) which occurs for bodies in relative motion.  We show that this $\mathpzc{PT}$-symmetry is a consequence of a universal frequency-momentum mirror symmetry observed under the relativistic Lorentz transformations, which was surprisingly overlooked so far. We analyze the case of metallic media in relative motion and show that the spectral singularity occurs because of the perfect coupling between positive and negative frequency surface plasmon polaritons. This is fundamentally different from the case of the balance between spatially distributed gain and loss profiles known in conventional $\mathpzc{PT}$-symmetric systems. These $\mathpzc{PT}$-symmetric spectral singularities are manifested at the transition between stable regions (loss dominant regime) and region of instabilities (gain dominant regime) in the dispersion of a moving system. Our work explains the underlying cause of a giant enhancement in all phenomena related to vacuum and thermal fluctuations in moving media eg.: vacuum forces and radiative heat transfer. We show that the giant enhancement is caused by the universal phenomena of coupling between negative and positive frequencies in the near-field and therefore can be used to explain similar effects in acoustic systems \cite{shi2016accessing}, hydrodynamic flows \cite{nath2016pure} and experiments on Coulomb drag \cite{nandi2012excitoncdrag}. 

%we show that $\mathpzc{PT}$-symmetry is achieved at a critical frequency-momentum condition in moving metal-insulator-metal (MIM) structure which results in a spectral singularity.  In section~II, we derive the condition of frequency-mirror-symmetry at which the frequency of a mode, as observed in a moving frame of reference, flips its sign while conserving its momentum. In section~III, we show that this frequency-mirror-symmetry manifests as negative frequency response of a moving metallic slab to a surface plasmon at positive frequency. In section~IV, we show that the coincidence of positive and negative frequency response arising from the frequency-mirror-symmetry leads to the $\mathpzc{PT}$-symmetry condition. We derive the condition of $\mathpzc{PT}$-symmetry (which we call the $\mathpzc{PT}$-symmetry line)  in the $\omega-k$ plane based on the frequency-mirror-symmetry condition. Finally in section~V, we show that as a propagating mode approaches the $\mathpzc{PT}$-symmetry line, its $Q$-factor of resonance approaches infinity, resulting in zero-width resonance or spectral singularity. 

\section{Frequency-momentum mirror symmetry}

In this section, we show the existence of a frequency-momentum mirror symmetry in the Lorentz transformation laws. The time-dependent electromagnetic field solutions to Maxwell's equations are real variables. Thus the spectral decomposition of modes necessarily consist of positive and negative frequencies which are complex conjugates of each other. For the dispersion relation in the $\omega-k$ plane, this implies that positive frequency branches are necessarily accompanied by a symmetric negative frequency branch.  Under stationary conditions, the positive frequency components alone contain all the physics in the system, and therefore it generally suffices to restrict our analysis to the positive frequencies. However, both the positive and negative frequency solutions become relevant when there is a relative translatory motion in the system. This is because the Doppler shifts are velocity dependent causing the symmetry between forward/backward traveling waves and positive/negative frequencies to be broken.

    \begin{figure}
		\centering
		\includegraphics[width=1\linewidth]{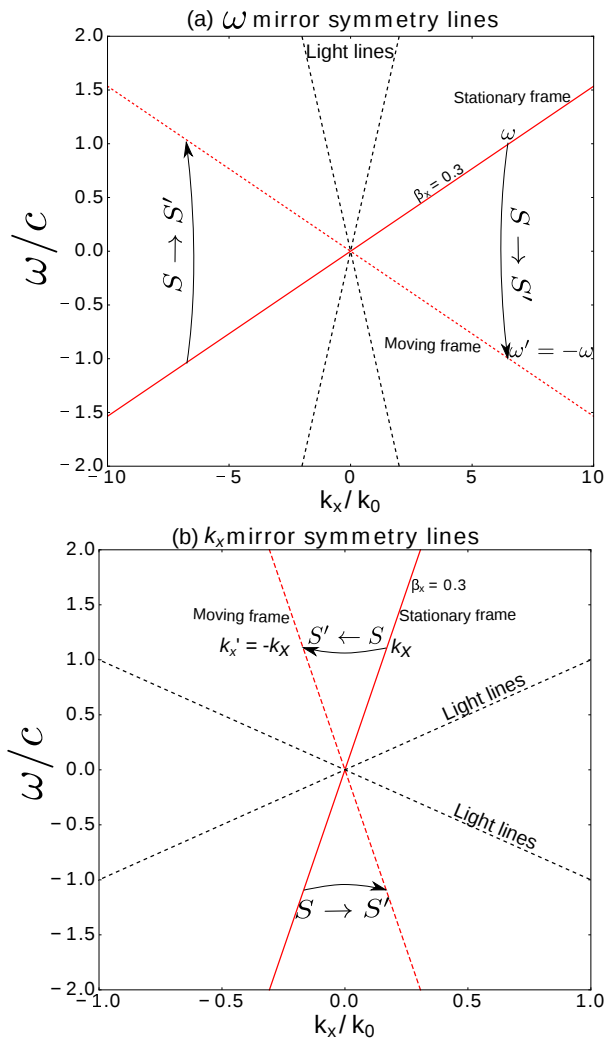}
		%\caption{(a) Frequency-symmetry lines (b) Momentum-symmetry lines. In (a) two frequency-symmetry lines are shown for $\beta_x=0.3$(red) and $\beta_x=0.12$(cyan). The Lorentz transformed lines observed in moving frame of reference $S'$ are shown as dashed line.}
		\caption{(a) Shows the frequency-mirror-symmetry condition for $\beta_x=0.3$. The equation of the line satisfying frequency-mirror symmetry condition (equation(\ref{eq:kx_symmetry})) in frame $S'$ is shown by the solid red line, while the corresponding Lorentz transformed line in the frame of reference $S'$ is shown by the dashed red line. It can be seen that the frequency ($\omega$) flips its sign while the momentum $k_x$ is conserved. (b) shows the momentum-symmetry-condition (equation(\ref{eq:omega_symmetry})), where the solid red line is the momentum-mirror-symmetry line in the $S$ frame of reference which transforms to the dashed red line the moving frame of reference $S'$. It can be seen that the momentum ($k_x$) changes its sign while frequency $\omega$ is conserved.}
		\label{fig:mirror_symmetries}
	\end{figure}
    
	The transformation of frequency ($\omega$) and momentum ($k_x$) from a stationary frame of reference $S$ to an inertial frame of reference $S'$, under the relativistic Doppler shift  is governed by \cite{kong1975theory},
	\begin{eqnarray}
	\label{eq:kx_transformation} k_x' &= \gamma \left( k_x - \beta_x\frac{\omega}{c} \right)\\
	\label{eq:omega_transformation} \omega'& =  \gamma\left(\omega - k_x\beta_x  c \right)
	\end{eqnarray}
	where $\omega'$ and $k_x'$ is the frequency and the propagation constant (respectively), as seen in the transformed frame of reference $S'$ moving with a velocity $v_{motion}$; $\beta_x=v_{motion}/c$ is the normalized velocity of translation, and $\gamma = 1/(1-\beta_x^2)^{1/2}$. $c$ is the velocity of light in vacuum. For simplicity, we have considered translatory motion along the $x$-axis only. The central theme of this paper is the unique relativistic transformation from positive frequencies to an equal and opposite frequency given by
	\begin{equation}
	\omega' = - \omega
	\label{eq:omega_negative_symmetry}
	\end{equation}
Note that the momentum of waves is invariant to this transformation and is conserved,
	\begin{equation}
	k_x'=k_x
	\label{eq:omega_negative_symmetry_kx}
	\end{equation}
	%When we consider the special case of relativistic transformation  such that momentum ($k_x$) in frame $S$ transforms to the same value in another inertial frame $S'$ (i.e. $k_x'=k_x$), from eq(\ref{eq:kx_transformation}), we get an equation of a line in the $\omega-k_x$ plane,
	This unique transformation is satisfied by the equation of a line
	\begin{equation}
	k_x = \frac{\gamma\beta_x}{c\left(\gamma-1\right)}\omega
	\label{eq:kx_symmetry}
	\end{equation}
	We call this as the frequency-mirror-symmetry condition, on which the frequency flips its sign in a transformed frame of reference while maintaining its momentum.

On similar lines, it can be shown that another special relativistic transformation exists which maps the wave momentum in the stationary to an equal and opposite momentum in the moving frame i.e. 
	\begin{equation}
	k_x' = -k_x
	\label{eq:kx_negative_symmetry}
	\end{equation}
This is satisfied on the line
	\begin{equation}
	\omega = \frac{\gamma\beta_x c}{\gamma-1}k_x
	\label{eq:omega_symmetry}
	\end{equation}
Note, that in this case, the frequency is invariant to the Lorentz transformation
	\begin{equation}
	\omega'=\omega
	\label{eq:kx_negative_symmetry_omega}
	\end{equation}
	
We call this as the momentum-mirror-symmetry condition.

Thus, the fundamental Lorentz transformation equations (\ref{eq:kx_transformation}),(\ref{eq:omega_transformation}) have universal symmetry properties in the $\omega-k_x$ plane such that for a given velocity $v_{motion}$, there exists,  % a locus of $(\omega,k_x)$ in $S$ on which 
	\begin{enumerate}
		\item A line given by equation~(\ref{eq:kx_symmetry}), on which the momentum is conserved ($k_x'=k_x$) while the frequency flips its sign (frequency-mirror-symmetry $\omega'=-\omega)$. 
		\item A line given by equation~(\ref{eq:omega_symmetry}), on which the frequency is conserved ($\omega'=\omega$) while the momentum flips its sign (momentum-mirror symmetry, $k_x'=-k_x$).
	\end{enumerate}
	%(1) the momentum is conserved ($k_x'=k_x$) while frequency changes sign with same magnitude(frequency-mirror-symmetry $\omega'=-\omega)$; (2) The frequency is conserved $\omega'=\omega$ while the momentum changes sign with same magnitude (momentum mirror symmetry, $k_x'=-k_x$).  

	These two symmetry conditions in the $\omega-k_x$ plane are shown in Fig.~\ref{fig:mirror_symmetries}, panel (a) and (b) respectively. The dashed black lines in the figure represent the light lines. In Fig.~\ref{fig:mirror_symmetries}(a), the solid red line represents the equation~(\ref{eq:kx_symmetry}) for $\beta_x=0.3$, in the stationary frame of reference denoted by $S$. The line then transforms to the dashed line in the moving frame of reference denoted by $S'$. 
	%In Fig.~\ref{fig:mirror_symmetries}(a), the solid red and cyan lines represents the equation~(\ref{eq:kx_symmetry}) for $\beta_x=0.3$ and $\beta_x=0.12$ respectively, in $S$ frame of reference. These lines then transform to the dashed lines of the corresponding colour in the moving frame of reference $S'$. The locus of the transformation of a point on the line is shown by the coloured line, with its colour mapped to the value of $\beta_x$. Same colored point on the line indicate mirror symmetry pairs. 
	It can be seen that a positive frequency mode satisfying equation~(\ref{eq:kx_symmetry}) transforms to its negative frequency counterpart while the momentum is invariant, i.e. it exhibits  frequency-mirror-symmetry. Similarly, Fig.~\ref{fig:mirror_symmetries}(b) shows the transformation of a momentum-mirror symmetry line. It should be noted that while the momentum-mirror symmetry line lies inside the light line, the frequency-mirror-symmetry condition can be satisfied only outside the light line, i.e. when the phase velocity is lower than the velocity of light. 
	
	The frequency-mirror-symmetry condition with its flipped frequency and invariant momentum (shown in Fig.~\ref{fig:mirror_symmetries}(a)) is of particular interest, because it enables the observation of the negative frequency electromagnetic response of a medium at positive frequencies. Thus, an EM mode at $(\omega,k_x)$ on the frequency-mirror-symmetry line, will be transformed to $(-\omega,k_x)$ in a moving medium, and consequently its negative frequency response will be observed in the stationary frame of reference.

	%In the next section, we show that this property results in observation of complex conjugate response at frequency-mirror-symmetry point in moving media. Later, we will show that this mirror-symmetry condition is the root of $\mathpzc{PT}$-symmetric spectral singularity in moving medium. 

	\section{Negative frequency response at positive frequencies}
Our mirror-symmetry arguments are completely general and apply in the relativistic case. Here, we show a practical scenario where this symmetry is manifested. First,we provide a physical interpretation of negative frequency modes followed by the role of the universal symmetry described above. The electromagnetic properties of a metal slab in relative motion with a velocity $v_{motion}$, as observed in the stationary lab frame of reference are governed by the Lorentz transformation of constitutive relations. However at low velocities ($\left(v_x/c\right)^2\ll 1$) under the First Order Lorentz transformation (FOLT) limit, the dielectric response of the slab as seen from the stationary lab frame is $\epsilon_r\approx\epsilon_r'(\omega')$ \cite{kong1975theory}, where $\epsilon_r'$ is the dielectric response of the metal and $\omega'$ is the frequency as seen in its proper frame of reference. $\omega'$ is obtained by transforming $\omega$ as per equation~(\ref{eq:omega_transformation}), which simplifies to $\omega'=\omega-k_x v_{motion}$ under FOLT limit. The dielectric response as seen in the lab frame of reference is then given by 
	\begin{equation}
	\epsilon_r=\epsilon_r(\omega-k_x v_{motion})
	\label{eq:epsilon_lab_moving}
	\end{equation}
	
	\begin{figure}
		\includegraphics[width=1\linewidth]{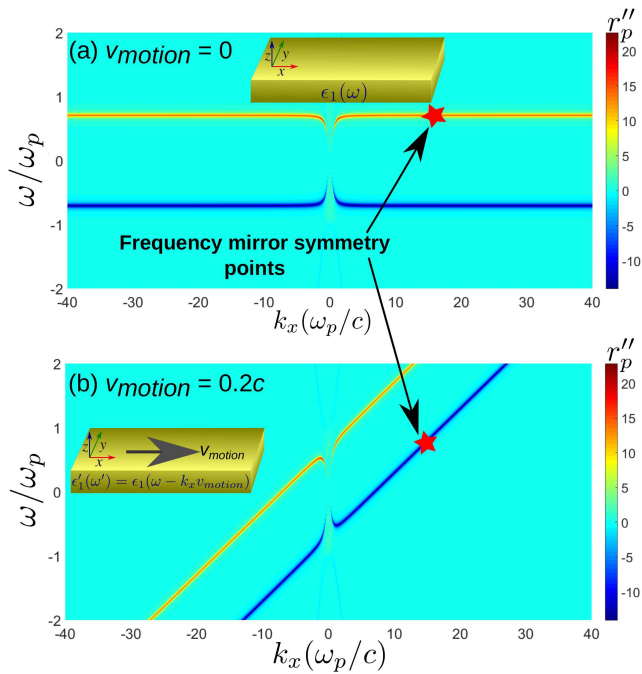}
		\caption{Imaginary part of reflection coefficient for (a) moving with velocity $v_x=0.1c$ and (b) stationary metallic slab in the four quadrants of $\omega-k_x$ plane. p-polarized plasmonic mode is considered such that $r_p = \frac{\epsilon_1 k_{z1}-\epsilon_2 k_{z2}}{\epsilon_1 k_{z1} +\epsilon_2 k_{z2}}$. Here, $k_{z1}=\sqrt{\epsilon_1 k_0^2-k_x^2}$, $k_{z2}=\sqrt{\epsilon_2 k_0^2-k_x^2}$ and dielectric response is governed by Drude model $\epsilon_1 (\omega)=1-\frac{\omega_p^2}{\omega^2+i\Gamma\omega}$. It can be seen that peak in $r_p^{\prime\prime}$ follows the dispersion curve of surface plasmon polaritons. The $r_p''$ is proportional to the normal component of the Poynting vector in the evanescent plasmonic wave, and is representative of the loss in the slab. For a lossy medium, $r_p''$ is positive for positive frequencies and negative for negative frequencies as depicted in panel (a). For a moving slab, in the region where $k_x v_{motion}>\omega$, the negative frequency mode (represented by the blue peak in $r_p''$) is dragged into the positive frequency quadrant, which exhibits gain at positive frequencies. Therefore in a moving slab there are two forward propagating modes in the positive frequency region; one is the positive frequency lossy mode represented by positive peaks in $r_p''$, while the other is the gain mode with negative $r_p''$ in the positive frequency region.}
		\label{fig:reflection_moving_MIM}
	\end{figure}
	It can be seen that the dielectric response of a moving metal slab is not just dependent on frequency, but also on the propagation constant and the velocity of motion. As a consequence, an incident wave of frequency $\omega$ and propagation constant $k_x$ will observe a negative frequency dielectric response of the metal when,
	\begin{align}
	\label{eq:chernkov_condition} \omega'  =  \omega & - k_x v_{motion} < 0 \\
	\label{eq:chernkov_condition_2}	 v_{motion}  &> v_p 
	\end{align}
	This is the Cherenkov condition at which the velocity of motion is greater than the phase velocity ($v_p=\omega/k_x$) of the wave \cite{erenkov1937visible,ginzburg1996radiation}. 
	
	The condition of $v_{motion}>v_p$ can be physically satisfied only when $v_p\ll c$. This requires the momentum to be larger than the free space wavevector $k_{x} > k_{0}=\omega/c$ causing the waves to decay in vacuum. Thus the negative frequency transformation for metallic slab in motion can occur for near-field evanescent waves. We now analyze the reflection properties of such evanescent waves at the vacuum-metal interface. The origin of such evanescent waves could be quantum emitters or another stationary slab in the near-field of the moving slab. The normal component of Poynting vector ($S_z$) of an evanescent wave absorbed at the metal interface is proportional to the imaginary component of the reflection coefficient ($r_p''$) \cite{guo2014singular}. 
		\begin{equation}
	S_z \propto r_p''
	\label{eq:poynting_z}
	\end{equation}
Therefore the reflection coefficient of a semi-infinite Drude metal ($\epsilon_r=1-\omega_p^2/(\omega^2+i\omega\Gamma)$) slab sheds light on the absorption and amplification characteristics of the medium. Note that the tangential boundary conditions and hence the Snell's reflection law at moving media interface is valid when the motion is in the plane of the interface \cite{kong1975theory} [see Appendix~A]. 
	
The sign of the imaginary component of the reflection coefficient ($r_p''$) is representative of the loss in the metal slab and it's peak follows the dispersion curves of a surface plasmon polariton (SPP). For any lossy metal, $r_p''$ is positive. However, this is strictly true only at positive frequencies. At negative frequencies, the dielectric response and the reflection coefficient is the complex conjugate of its respective positive frequency values ($\epsilon(-\omega)=\epsilon^{*}(\omega)$). 

Fig.~\ref{fig:reflection_moving_MIM}(a) shows the dispersion of the p-polarized plasmonic mode in the $\omega-k_x$ plane. It can be seen that the SPPs have positive peaks in $r_p''$ for positive frequencies ($\omega>0$ region), and negative peaks for negative frequencies ($\omega<0$ region). The negative frequency region actually corresponds to the complex conjugate part of the electromagnetic field solution. The mode solutions have phase fronts which are forward propagating when $k_x$ and $\omega$ have the same sign and when they have opposite signs the mode is backward propagating. Therefore the complete representation of a forward propagating mode includes the first and third quadrant solution, while that of a backward propagating mode includes second and fourth quadrant. It should also be noted that in the stationary case, the dispersion characteristics are symmetric for the forward ($k_x>0$) and the backward propagation ($k_x<0$) as well as positive and negative frequencies.
	
Motion of the slab breaks the symmetry of the dispersion relation since the Lorentz transformation of frequency and momentum (or equivalently the fields) is velocity dependent. In the extreme case, when the slab is moving with velocity above the Cherenkov limit, the negative frequency mode from the fourth quadrant is dragged into the positive frequency region. Note the phase of the wave is an invariant of motion and the condition for real EM fields holds true in all reference frames. The symmetry breaking in the dispersion relation due to motion is shown in Fig.~\ref{fig:reflection_moving_MIM}(b). The negative frequency mode which is dragged into the positive frequency region has negative $r_p''$, implying gain characteristics in the positive frequency domain. A slab moving above the Cherenkov limit has two forward propagating modes and no backward propagating mode. One of the forward propagating modes is an ordinary lossy mode (shown by red peak in $r_p''$), while the other mode is amplified  (growing mode shown by blue peak in $r_p''$). The motion of the dielectric slab results in the violation of time reversal symmetry. 
	\begin{equation}
	\omega\left(k_x\right)\neq\omega\left(-k_x\right)
	\label{eq:non_reciprocal}
	\end{equation}
We would like to emphasize that the above argument is valid even at relativistic velocities, even without FOLT approximations. 

Note the existence of a special frequency-mirror-symmetry when the observed response of the moving slab is the exact complex conjugate of its stationary value. This is shown by the starred point in the $\omega-k_x$ plane. 
While the dielectric response at this frequency-mirror-symmetry point is $\epsilon_r(\omega, k_x)$ for a stationary slab, it is $\epsilon_r(-\omega,k_x)=\epsilon_r^*(\omega,k_x)$ for the moving slab. If we now consider two identical parallel slabs (see Fig. 3), one in relative motion to the other, our analysis shows the existence of a unique velocity at which the dielectric response of the moving slab is the complex conjugate of the stationary slab. This occurs for a specific frequency and momentum wavevector dictated by two conditions - the dispersion relation of surface waves on the slab and the frequency-momentum mirror symmetry condition.

	\section{$\mathpzc{PT}$-symmetric resonance in a moving MIM waveguide}
		
	\begin{figure}[h]
		\centering
		\includegraphics[width=1\linewidth]{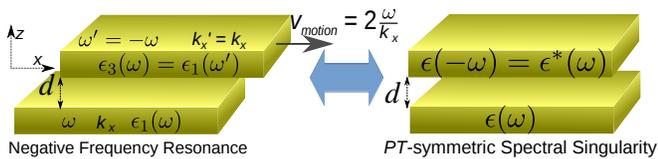}
		\caption{Metal slabs separated by a distance $d$ at relative motion interact via near field evanescent waves. When the velocity of motion is greater than the Cherenkov velocity ($v_{motion}>\omega/k_x$), the interaction is between positive frequencies of stationary slab and negative frequencies of the moving slab. This can result in a perfect coupling between positive and negative frequency modes which we call as a negative frequency resonance. When the velocity of motion is twice the Cherenkov limit ($v_{motion}=2\omega/k_x$), the dielectric response of the two slabs become complex conjugate pairs, resulting in $\mathpzc{PT}$-symmetric spectral singularity. The $\mathpzc{PT}$-symmetry is achieved as a consequence of the negative frequency-mirror-symmetry.}
		\label{fig:MIM_moving_slab}
	\end{figure}

Here, we show how the relativistic negative frequency-mirror-symmetry is connected to the achievement of parity-time symmetry in a moving system. Note our work in this section is connected to the mirror-symmetry condition and not the instabilities or spontaneous $\mathpzc{PT}$-symmetry breaking in moving media \cite{silveirinha2014spontaneous}. If we place a stationary and a moving metal slab close enough to allow evanescent wave interactions, they form a metal-insulator-metal (MIM) waveguide structure. In this structure, the stationary slab will have a dielectric response $\epsilon_1(\omega)$ and the moving slab will exhibit dielectric response of $\epsilon_3(\omega)=\epsilon_1(\omega')$. This is shown in Fig.~\ref{fig:MIM_moving_slab}. The separation between the slabs is $d$. The overall dielectric distribution of the waveguide as a function of $z$ is then written as,

    %\onecolumngrid
		
		\begin{figure*}
			\includegraphics[width=1.0\linewidth]{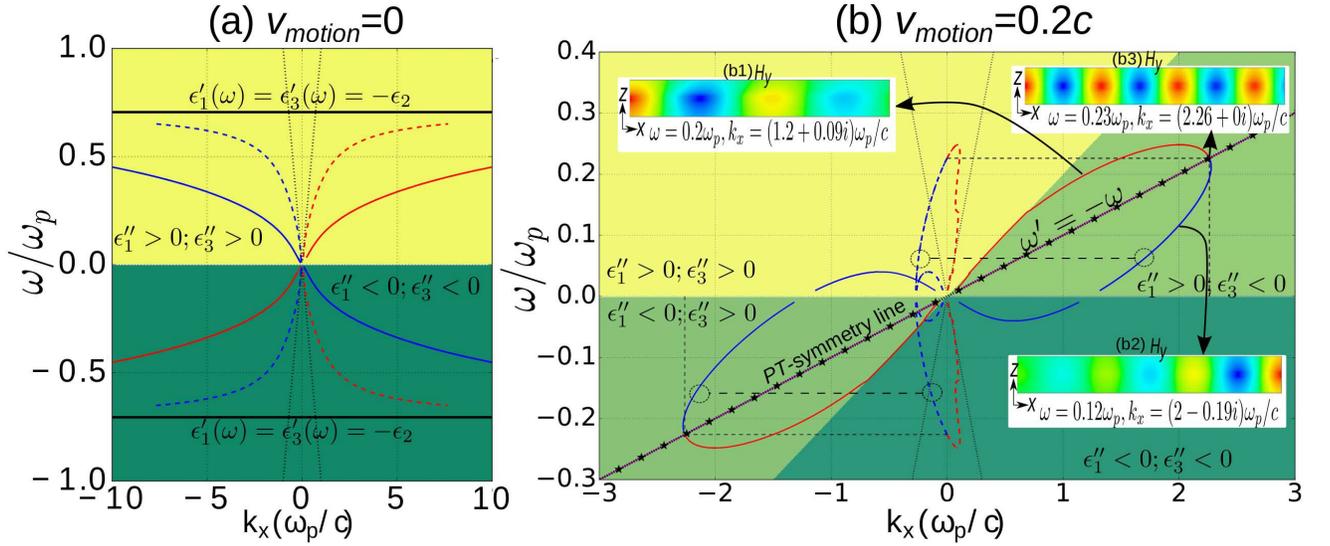}\\
			\caption{Dispersion curves for plasmonic MIM waveguide in four quadrants of $\omega-k_x$ plane for (a) stationary slabs (b) slabs with relative velocity of $0.2c$. Solid colored lines represent the real part of propagation constant $k_x^\prime$ while the dash-dot lines represent the imaginary part $k_x^{\prime\prime}$ of the respective mode. Black dotted lines represent line cone. In (b), the dispersion curve lying below the $\mathpzc{PT}$-symmetry line is the negative frequency mode which is dragged into the first quadrant. This the is the gain mode. The gain and loss modes merge at the intersection of $\mathpzc{PT}$-symmetry line with imaginary component of propagation constant equal to zero at the intersection point. At this point, loss in stationary metal slab is exactly compensated by gain in the moving metal slab. All points on the dispersion curve below the $\mathpzc{PT}$-symmetry line have gain while all the points above have loss. Inset shows $H_z$ mode profile at loss (b1) , gain (b2) and $\mathpzc{PT}$-symmetric (b3) points. The magnetic field profile in the lossy region of the dispersion curve (b1) attenuates as it propagates along $x$ direction, and amplifies in the gain region (b2). On the $\mathpzc{PT}$-symmetry point of the dispersion curve, the field profiles propagates without any attenuation or gain. Drude metal with plasma frequency $\omega_p=10^{14} Hz$, collision frequency $\Gamma=0.05\omega_p$ is considered in the simulation. The separation between the slabs is $d=25$~nm.}
			\label{fig:dispersion_relation_MIM_moving_betax=0_d=25nm_epsilon2=1_Gamma=0pt05}
		\end{figure*}
		
		%\twocolumngrid 

	\begin{equation}
	\epsilon(z,\omega)=\begin{cases}
	\epsilon_1(\omega) &;  	z<-d/2\\
	\epsilon_2=1 	&; 	 -d/2< z<d/2	\\
	\epsilon_1(\omega') &; d/2<  z   
	\end{cases}
	\label{eq5_spp_moving}
	\end{equation}

The dielectric function of the waveguide is complex in the region $z<-d/2$ and $z>d/2$. To investigate the $\mathpzc{PT}$-symmetry properties of this complex dielectric system, we write the Hamiltonian formulation for a plane wave propagation ($e^{ik_x x}$) along $\hat{x}$ direction  \cite{kong1975theory, alaeian2014non}
		
	\begin{equation}\label{Eq:EigenValue}
	%\hat{H}_{em} \textbf{$\Psi$}_{k_x}(y,z)=k_x^2   \textbf{$\Psi$}_{k_x} (y,z)
    \hat{H}_{em} \textbf{$\Psi$}_{k_x}(z)=k_x^2   \textbf{$\Psi$}_{k_x} (z)
	\end{equation}
	where eigenfunctions \textbf{$\Psi_{k_x}$} can be either $\vec{E}(y,z)$ or $\vec{H}(y,z)$ and $k_x^2$ is the eigenvalue. Assuming non-magnetic media and in the FOLT limit, $\hat{H}_{em}$ can be written as
	\begin{equation}\label{Eq:Hamiltonian}
	\hat{H}_{em}(z,\omega)=\textbf{$\nabla$}_t^2+\omega^2\epsilon_0\mu_0\epsilon(z,\omega),
	\end{equation}
in which $\textbf{$\nabla$}_t^2=\hat{x}\times \textbf{$\nabla$}^2$. Equation~(\ref{Eq:EigenValue}) will be $\mathpzc{PT}$-symmetric with real eigenvalues and unitary time-evolution if \textbf{$\Psi_{k_x}$} is eigenfunction of $\mathpzc{PT}$ operator and,	
	\begin{equation}
	[\hat{H}_{em},\mathpzc{PT}]=0
	\label{Eq:PT_H_commute}
	\end{equation}

	From the properties of $\mathpzc{P}$ and $\mathpzc{T}$ operators, it is straightforward to show that [see Appendix B]

	\begin{subequations}\label{Eq:H&P&T}
		\begin{equation}\label{Eq:H&P}
		\mathpzc{P}.\hat{H}_{em}(z,\omega)=\hat{H}_{em}(-z,\omega).\mathpzc{P}
		\end{equation}
		\begin{equation}\label{Eq:H&T}
		\mathpzc{T}.\hat{H}_{em}(z,\omega)=\hat{H}_{em}^*(z,\omega).\mathpzc{T}
		\end{equation}
	\end{subequations}

	\indent Combining Eq.~(\ref{Eq:H&P}) and (\ref{Eq:H&T}) together, we conclude that the Hamiltonian is $\mathpzc{PT}$-symmetric if $\hat{H}_{em}(z,\omega)=\hat{H}_{em}^*(-z,\omega)$. This means that the condition of $\mathpzc{PT}$-symmetry on dielectric function, as obtained from equation~(\ref{Eq:Hamiltonian}), is the well known condition \cite{principe2015supersymmetry}
	\begin{equation}
	\epsilon(-z,\omega)=\epsilon^*(z,\omega).
	\label{Eq:pt_condition_on_epsilon}
	\end{equation}
	
Using equation~(\ref{eq5_spp_moving}) and the fact that the imaginary part of dielectric response is an odd function of frequency \cite{landau2013electrodynamics}, the condition for $\mathpzc{PT}$-symmetry in the moving system translates to,
	\begin{equation}
	\epsilon_1\left(\omega\right)=\epsilon_1(-\omega')
	\label{eq:epsilon_pt_symmetry}
	\end{equation}
	
This condition is only met when $\omega'=-\omega$ for the same value of $k_x$ in stationary as well as moving frame of reference, i.e. on the frequency-mirror-symmetry line of equation~(\ref{eq:kx_symmetry}). This is a unique case where the system response is  $\mathpzc{PT}$-symmetric only for a specific electromagnetic mode. Our moving slab systems does not possess time-reversal-symmetry or parity-symmetry individually for any mode.  In the FOLT limit, the frequency-mirror-symmetry line simplifies to [see Appendix C], 
	\begin{equation}
	k_x=2\frac{\omega}{v_{motion}}
	\label{eq:pt-symmetry_line}
	\end{equation}

	We will henceforth refer to this line as the $\mathpzc{PT}$-symmetry line along which the Hamiltonian (equation~(\ref{Eq:Hamiltonian})) is $\mathpzc{PT}$-symmetric or $[\mathpzc{PT},\hat{H}_{em}]=0$. Note that this condition is independent of the separation $d$ but the spectral singularity depends on the gap distance. A mode of the system on the $\mathpzc{PT}$-symmetry line will not undergo attenuation or amplification, because at this condition the loss in the stationary slab is perfectly balanced by the gain in the moving slab.  

\section{Cherenkov amplification}

We emphasize that the parametric amplification of vacuum fluctuations is well-known for the phenomenon of vacuum friction which occurs for bodies in relative motion \cite{pendry1997shearing}. The parametric nature arises since the evanescent wave on reflection is amplified without change in frequency or momentum. Similarly, growing electromagnetic waves and instabilities in moving plasmas and their connection to negative energy waves have been well-studied \cite{nezlin1976negative}. However, the role of the frequency-momentum mirror symmetry condition in Lorentz transformations, the perfect coupling of positive and negative frequencies in the near-field and $\mathpzc{PT}$-symmetric spectral singularity has never been pointed out till date.

We note that gain in the moving system arises from Cherenkov amplification also known as the anomalous Doppler effect fundamentally different from conventional $\mathpzc{PT}$-symmetric systems.  These growing waves in moving media can be seeded by vacuum fluctuations. We note that unlike the classical Cherenkov radiation where charged particles are necessary, this effect only requires the motion of neutral polarizable particles or harmonic oscillators with internal degrees of freedom. The classical dispersion relation of modes enters the classical thermal fluctuations and quantum vacuum fluctuations through the fluctuation-dissipation theorem (FDT) \cite{levin1980contribution}. 

\begin{equation}
\langle E_j(\vec{r},\omega)E_k^*(\vec{r}',\omega)\rangle = \frac{\omega}{2\pi c^2\epsilon_0} \Theta(\omega,T) \mathrm{Im} \left\{G_{jk}^E(\vec{r},\vec{r}';\omega)\right\}
\label{eq:FDT1}
\end{equation}
where $\hat{j}$ and $\hat{k}$ represent the three spatial orthogonal coordinates ($\hat{j},\hat{k}\in\{\hat{x},\hat{y},\hat{z}\}$). $\Theta(\omega,T)$ is the energy of a quantum oscillator at equilibrium, given by $\Theta(\omega,T) = \hbar\omega/2 + \hbar\omega/(e^{\hbar\omega/(K_BT)}-1)$, and $G_{jk}$ is an element of the electric field Green's tensor.  
We note that the FDT formalism ensures that classical mode structure affects the noise properties and effects such as Casimir forces \cite{milton2001casimir,philbin2009casimir}, near-field heat transfer \cite{liu2015enhanced}  and vacuum friction \cite{pendry1997shearing}.  

\section{$\mathpzc{PT}$-symmetric spectral singularity}
	
In this final section, we show the spectral singularity associated with $\mathpzc{PT}$-symmetric systems is manifested in the perfect coupling of positive and negative frequency branches in moving plasmonic media. This perfect coupling occurs at a critical velocity and gap distance when negative frequency mirror symmetry is achieved for surface wave solutions. We call this a negative frequency resonance. Note that, even though $\mathpzc{PT}$-symmetry is satisfied on all points of the $\mathpzc{PT}$-symmetry line, not all $(\omega,k_x)$  on the line are valid waveguide modes. This is because a mode has to satisfy additional boundary conditions at the interfaces, which also makes the solution gap-size ($d$) dependent. To get the precise location of a mode on the $\mathpzc{PT}$-symmetry line, we compute the full dispersion curve for an MIM waveguide with one moving slab, by solving the dispersion relation,
	\begin{equation}
	e^{-2i k_{z2}d}=\frac{ k_{z2}/\epsilon_2 + k_{z1}/\epsilon_1}{ k_{z2}/\epsilon_2 - k_{z1}/\epsilon_1}\frac{ k_{z2}/\epsilon_2 + k_{z3}/\epsilon_3}{ k_{z2}/\epsilon_2 - k_{z3}/\epsilon_3}
	\label{eq:disp_relation_MIM}
	\end{equation} 
	in the full $\omega-k_x$ plane. Here, $k_{zj}=\sqrt{(\epsilon_j k_0^2-k_x^2)}$, $j\in \{1,2,3\}$; $\epsilon_1$ is the Drude dielectric response, $\epsilon_2=1$ (or a constant)  and $\epsilon_3=\epsilon_1(\omega-k_x v_{motion})$. We consider p-polarized ($H_y\neq 0$) wave propagation as it alone supports plasmonic modes.

Fig.~\ref{fig:dispersion_relation_MIM_moving_betax=0_d=25nm_epsilon2=1_Gamma=0pt05}  contrasts the computed dispersion curves of the MIM waveguide with (a) stationary slabs and (b) one slab moving with $v_{motion}=0.2c$. Background color in the plot indicates the sign of the imaginary part of the dielectric response of the two slabs ($\epsilon_1$ and $\epsilon_3$). In Fig.~\ref{fig:dispersion_relation_MIM_moving_betax=0_d=25nm_epsilon2=1_Gamma=0pt05}a, dark yellow background represents the region where $\epsilon_1^{\prime\prime}$ and $\epsilon_3^{\prime\prime}$ are positive, while dark green represents the region where both the constants are negative. For a lossy medium, $\epsilon^{\prime\prime}$ is positive for positive frequencies and negative for negative frequencies. Thus both slabs ($\epsilon_1$ and $\epsilon_3$) are lossy in the stationary case. 

When one of the slabs starts moving, its dielectric response transforms according to $\epsilon(\omega-k_x v_{motion})$. Above the Cherenkov limit of $v_{motion}>\omega/k_x$, negative frequency characteristics are dragged into the positive frequency region as shown by the overlap of light green and light yellow region in Fig.~\ref{fig:dispersion_relation_MIM_moving_betax=0_d=25nm_epsilon2=1_Gamma=0pt05}b. In this overlap region medium-1 is lossy, while medium-3 exhibits amplifying characteristics. The $\mathpzc{PT}$-symmetry line lies in this overlap region and forms the diagonal to the rhombus formed between the region $\epsilon_1<-\epsilon_2$ and $\epsilon_3<-\epsilon_2$. The condition $\epsilon_1=\epsilon_3^*$ is satisfied at all points on $\mathpzc{PT}$-symmetry line which implies the moving slab dielectric response is the complex conjugate of the stationary slab.
	
\begin{figure}[t]
	\centering
	\includegraphics[width=1\linewidth]{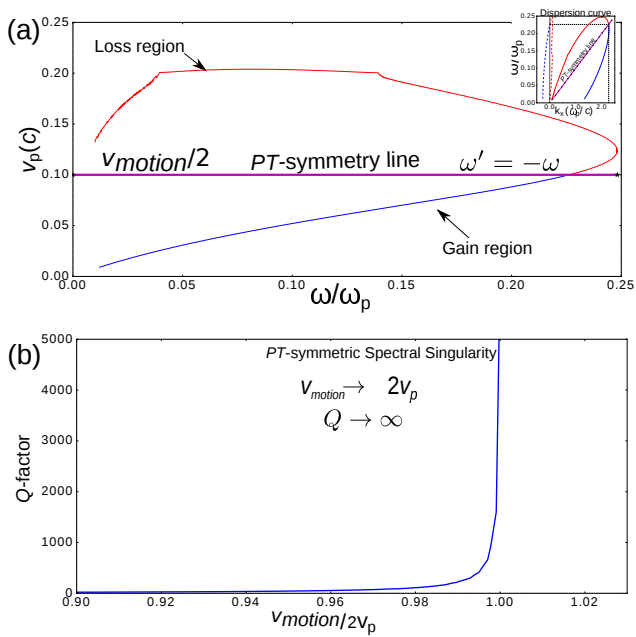}
	\caption{(a) Phase velocity along the dispersion curve as function of frequency when $v_{motion}=0.2c$. The gain and loss regions of the dispersion curve are separated by the $\mathpzc{PT}$-symmetry line with $v_{motion}/2$. The mode intersects the $\mathpzc{PT}$-symmetry line when $v_{motion}=2v_p$. (b) $Q$-factor of the resonance as a function of the ratio $v_{motion}/2v_p$. The mode exhibits spectral singularity as $v_{motion}\rightarrow 2v_p$, at the intersection of the $\mathpzc{PT}$-symmetry line and dispersion curve.}
	\label{fig:vp_Q-factor_MIM_moving_betax=0pt2_d_25nm_epsilion2_1_Gamma_0pt05_first_quad}
\end{figure}

In Fig.~\ref{fig:dispersion_relation_MIM_moving_betax=0_d=25nm_epsilon2=1_Gamma=0pt05}, the real component of propagation constant ($k_x^\prime$) is represented by solid lines and the imaginary component ($k_x^{\prime\prime}$) is shown by the dashed line of corresponding color. For the stationary MIM waveguide (Fig.~\ref{fig:dispersion_relation_MIM_moving_betax=0_d=25nm_epsilon2=1_Gamma=0pt05}a), both positive as well as negative momentum waves attenuate as they propagate in either direction, as indicated by the same sign of $k_x^\prime$ and $k_x^{\prime\prime}$ for all positive frequencies ( The signs are opposite in the complex conjugate region of negative frequencies). However when one slab is moving, we notice two unique modes in first quadrant, one is a lossy mode ($k_x^\prime$ and $k_x^{\prime\prime}$ have same sign) and another is a mode exhibiting gain ($k_x^\prime$ and $k_x^{\prime\prime}$ have opposite signs). These two modes are shown by red and blue colors in Fig.~\ref{fig:dispersion_relation_MIM_moving_betax=0_d=25nm_epsilon2=1_Gamma=0pt05}b, respectively. The gain mode in first quadrant arises form the negative frequency component of the backward propagating mode which is dragged to the positive frequency region from the fourth quadrant. The lossy and amplified modes converge and meet on a point on $\mathpzc{PT}$-symmetric line (shown by magenta colored line with star marker). We emphasize that the propagation constant at this point of intersection is purely real, $k_x^{\prime\prime}=0$. At this point on the dispersion curve, the $\mathpzc{PT}$-symmetry is achieved and the wave propagates without any attenuation. All points on the dispersion curve above the $\mathpzc{PT}$-symmetry line are lossy, while those below exhibit gain as shown by the $H_y$ mode profile in inset. It can also be seen that, in contrast to the stationary case, the dispersion diagrams becomes unsymmetrical $\omega(k_x)\neq\omega(-k_x)$.

	The dispersion curve and the $\mathpzc{PT}$-symmetry line intersect when the phase velocity $(v_p)$ of the mode is equal to half the slab velocity. 	
	Fig.~\ref{fig:vp_Q-factor_MIM_moving_betax=0pt2_d_25nm_epsilion2_1_Gamma_0pt05_first_quad}(a) shows the phase velocity of points along the dispersion curve when $v_{motion}=0.2c$. The corresponding dispersion curve in the first quadrant of the $\omega-k_x$~plane is shown in the inset. All the points on dispersion curve with phase velocity higher than $v_{motion}/2$ are lossy, while those with lower phase velocity exhibit gain. Thus the Cherenkov limit for amplification for MIM waveguide with one moving slab, is modified to
	\begin{equation}
	v_{motion}>2v_p	
	\label{eq:modified_cheernkov_limit}
	\end{equation}
	The more stringent condition for amplification arises from the fact that the gain in the moving slab has to compensate for the loss in stationary slab, and therefore to achieve net gain the slab velocity has to be twice the conventional Cherenkov limit. The $\mathpzc{PT}$-symmetry condition lies at the boundary of stability (loss dominant regime) and instability (gain dominant regime).

	A mode at the $\mathpzc{PT}$-symmetry condition ($v_{motion}=2v_p$) exhibits zero-width resonance, as depicted by the $Q$-factor of resonance in Fig.~\ref{fig:vp_Q-factor_MIM_moving_betax=0pt2_d_25nm_epsilion2_1_Gamma_0pt05_first_quad}(b). The $Q$-factor is defined as a ratio  $k_x'/2k_x''$ [see Appendix D]. It can be seen that at the $\mathpzc{PT}$-symmetry condition, the $Q$-factor tends to infinity, indicating zero-width of resonance or spectral singularity. Note that non-equilibrium phenomena such as radiative heat transfer and vacuum friction will exhibit a giant enhancement when the velocity and gap size is tuned to achieve this resonance \cite{guo2014giant} . Our future work will focus on theoretical work beyond linear response theory to regularize the fluctuations near this spectral singularity.

	\section{Conclusion}	
In this paper, we have shown the existence of a universal frequency and momentum mirror symmetry conditions in relativistic Lorentz transformations. We have shown that frequency-mirror-symmetry is the fundamental origin of the $\mathpzc{PT}$-symmetry condition in the case of metallic slabs in relative motion. We show that the $\mathpzc{PT}$-symmetry condition is achieved only on a line which satisfy frequency-mirror-symmetry. Our work provides a clear connection between negative frequency resonances and $\mathpzc{PT}$-symmetric spectral singularity  in moving media. We have considered two metallic slabs in motion to show how the spectral singularity results from perfect coupling of positive and negative frequency surface plasmon polariton branches. Our work on the coupling of negative and positive frequencies in the near-field is a universal phenomenon and can lead to similar effects being discovered in acoustic systems \cite{shi2016accessing}, hydrodynamic flows \cite{nath2016pure} and experiments on Coulomb drag \cite{nandi2012excitoncdrag}.

	\bibliography{reference_moving_media_pt_symmetry}

	%\newpage
	
	\appendix
	\section{Boundary conditions in moving media interface}
	The generalized boundary conditions at moving media interface are derived in A. J. Kong's book \cite{kong1975theory}. For completeness, we restate these conditions and show that they simplify to stationary boundary conditions in our configuration. 
    
    At moving interface the electric and magnetic boundary conditions are not decoupled \cite{kong1975theory}
    \begin{align}
    \label{eq:appendix_a_1}&\hat{n}\times\left(\vec{E_1}-\vec{E_2}\right)-\left(\hat{n}\cdot\vec{v}_{motion}\right)(\vec{B_1}-\vec{B_2})=0 \\
    \label{eq:appendix_a_2}&\hat{n}\times\left(\vec{H_1}-\vec{H_2}\right)+\left(\hat{n}\cdot\vec{v}_{motion}\right)(\vec{D_1}-\vec{D_2})=\vec{J}_s
    \end{align}
where $\hat{n}$ is the normal to the interface. In our case the $v_{motion}$ is along the interface with $\hat{n}\perp\vec{v}$. Therefore the coupled terms with $\hat{n}\cdot\vec{v}$ have zero contribution and the boundary conditions simplify to,
 \begin{align}
    \label{eq:appendix_a_3}&\left(\vec{E}_{tangential1}-\vec{E}_{tangential2}\right)=0 \\
    \label{eq:appendix_a_4}&\left(\vec{H}_{tangential1}-\vec{H}_{tangential2}\right)=\vec{J}_s
    \end{align}
The boundary conditions on $\vec{D}$ and $\vec{B}$ derived from the divergence equations remain unaltered for moving media. Therefore the Snell's law is valid when the motion is in the plane of the interface. The reflection properties can then be computed by Lorentz transforming the material properties of moving medium to stationary frame of reference and applying the boundary conditions for stationary interface. Alternatively, the same reflection properties can be obtained by solving the boundary conditions in the proper frame of reference of the moving medium followed by Lorentz transformation of the solution to stationary frame of reference. 

%The plots in Fig.~\ref{fig:reflection_moving_MIM}(b) can be obtained by transforming the $\omega-k_x$ space of Fig.~\ref{fig:reflection_moving_MIM}(a) as per Lorentz transformation equation for $S'\rightarrow S$.

	\section{Parity and Time reversal operation}
	Parity reversal operation $\mathpzc{P}$ performs $r\rightarrow r'$ on the system. Therefore $\mathpzc{P}$ acting on the Hamiltonian of equation~(\ref{Eq:Hamiltonian}) gives, 
	\begin{align}
	&\mathpzc{P}.\hat{H}_{em}(z,\omega)=\mathpzc{P}.\left[\textbf{$\nabla$}_t^2+\omega^2\epsilon_0\mu_0\epsilon(z,\omega)\right]\\
	\implies & \mathpzc{P}.\hat{H}_{em}(z,\omega) = \! \left[(-\nabla_t)^2 + \omega^2\epsilon_0\mu_0\epsilon(-z,\omega) \right].\mathpzc{P}\\
	\implies & \mathpzc{P}.\hat{H}_{em}(z,\omega) = \hat{H}_{em}(-z,\omega).\mathpzc{P}
	\end{align}
	
	The time reversal operator $\mathpzc{T}$ is defined to reverse the direction of time, by performing $t\rightarrow -t $ and $i\rightarrow i^*$ on the system. Therefore $\mathpzc{T}$ acting on the Hamiltonian of equation~(\ref{Eq:Hamiltonian}) gives, 
	
	\begin{align}
	&\mathpzc{T}.\hat{H}_{em}(z,\omega) = \mathpzc{T}.\left[\textbf{$\nabla$}_t^2+\omega^2\epsilon_0\mu_0\epsilon(z,\omega)\right]\\
	\implies & \mathpzc{T}.\hat{H}_{em}(z,\omega) = \left[\nabla_t^2 + \omega^2\epsilon_0\mu_0\epsilon^*(-z,\omega) \right].\mathpzc{T}\\
	\implies & \mathpzc{T}.\hat{H}_{em}(z,\omega) = \hat{H}_{em}^*(z,\omega).\mathpzc{T}
	\end{align}

	\section{Frequency-mirror-symmetry in First Order Lorentz Transform (FOLT) limit}
	The frequency-mirror-symmetry line for a given $\beta_x$ is,
	
	\begin{equation}
	k_x = \frac{\gamma\beta_x}{c\left(\gamma-1\right)}\omega
	\label{eq:appendix_c_1}
	\end{equation}
	Substituting the value $\gamma=1/\sqrt{1-\beta_x^2}$, we get
	
	\begin{equation}
	k_x = \frac{\omega}{c}\frac{\beta_x}{1-\sqrt{1-\beta_x^2}}
	\label{eq:appendix_c_2}
	\end{equation}
	Using the binomial expansion,
	\begin{eqnarray}
	\left(1-\beta_x^2\right)^{1/2} = 1- \frac{1}{2}\beta_x^2 +\frac{1}{2}\frac{1}{2}\left(\frac{1}{2}-1\right)\beta_x^4 + \dots
	\label{eq:appendix_c_3}
	\end{eqnarray}
	
	In the first order Lorentz transform limit, $\beta_x^2 \ll 1$, and therefore we neglect the higher powers of $\beta_x$. Substituting equation~\ref{eq:appendix_c_3}) in equation~(\ref{eq:appendix_c_2}) and simplifying, we get 
	\begin{equation}
	k_x = 2\frac{\omega}{v_x}
	\label{eq:appendix_c_4}
	\end{equation}

	\section{Q-factor of a propagating wave resonance}
	The line width of resonance in a system is characterized by its $Q$-factor, defined as,
	
	\begin{equation}
	Q = \omega\frac{\textrm{Average~stored~Energy}(W)}{\textrm{Average~power~dissipated}(P)}
	\label{eq:appendix_b_1}
	\end{equation}
	The line width is inversely proportional to $Q$ value and  there is singularity in the spectrum when $Q\rightarrow \infty$. 
	
	The average stored energy in the system for time harmonic fields propagating in $x$ direction is,
	\begin{equation}
	W = \int_{vol} \vec{E}(x,t)\cdot \vec{E}^*(x,t) d(vol) + \int_{vol} \vec{H}(x,t)\cdot \vec{H}^*(x,t) d(vol)
	\label{eq:appendix_b_2}
	\end{equation}
	
	The electric and magnetic fields of a mode in one-dimensional propagation can be written as,
	
	\begin{align}
	\label{eq:appendix_b_3} \vec{E}(x,t)=\vec{E}e^{i(k_x' x -\omega t)}e^{-k_x'' x}\\
	\label{eq:appendix_b_4} \vec{H}(x,t)=\vec{H}e^{i(k_x' x -\omega t)}e^{-kx'' x}
	\end{align}
	where the complex propagation constant $k_x=k_x'+ik_x''$. From (\ref{eq:appendix_b_2}), (\ref{eq:appendix_b_3}) and (\ref{eq:appendix_b_4}), the average stored energy in the mode is,
	\begin{equation}
	W = \int_{vol}\vec{E}\cdot\vec{E}^* e^{-2k_x'' x}d(vol) + \int_{vol}\vec{H}\cdot\vec{H}^* e^{-2k_x'' x}d(vol)
	\label{eq:appendix_b_5}
	\end{equation}
	
	Average power dissipated by the mode as it propagates a distance $\partial x$ in time $\partial t$ is,
	\begin{equation}
	P = - \frac{\partial }{\partial t}(W)
	\label{eq:appendix_b_6}
	\end{equation}
	Substituting (\ref{eq:appendix_b_5}) in (\ref{eq:appendix_b_6}),
	\begin{equation}
	\begin{split}
	P = & \int_{vol}\vec{E}\cdot\vec{E}^* e^{-2k_x'' x} \left(2k_x''\frac{\partial x}{\partial t}\right)d(vol) \\
	+ &\int_{vol}\vec{H}\cdot\vec{H}^* e^{-2k_x'' x}\left(2k_x''\frac{\partial x}{\partial t}\right)d(vol)
	\end{split}
	\label{eq:appendix_b_7}
	\end{equation}
	Since a phase front ($k_x' x -\omega t = const$) travels $\partial x$ distance in time $\partial t$, we have 
	\begin{equation}
	\frac{\partial x}{\partial t} = \frac{\omega}{k_x'}
	\label{eq:appendix_b_8}
	\end{equation}
	
	From (\ref{eq:appendix_b_5}), (\ref{eq:appendix_b_7}) and (\ref{eq:appendix_b_8}), the average power dissipated is given by,
	\begin{equation}
	P= 2\frac{k_x''}{k_x'}\omega W
	\label{eq:appendix_b_9}
	\end{equation}
% 	The power dissipation is positive in a system with loss ($k_x''$ and $k_x'$ have same sign) and is negative in a system with gain ($k_x'$ and $k_x''$ have opposite sign). The line width of resonance will depend on the magnitude of power dissipated, irrespective of the sign of dissipated power.
From (\ref{eq:appendix_b_9}) and (\ref{eq:appendix_b_1}) we get the $Q$-factor of the propagating mode as,

	\begin{equation}
	Q = \frac{k_x'}{2 k_x''}
	\label{eq:appendix_b_10}
	\end{equation}

\end{document}